\documentclass[prl,amsmath,floats,floatfix, twocolumn,
superscriptaddress,nofootinbib,showpacs]{revtex4-1}

\usepackage{amsfonts}
\usepackage{amsmath}
\usepackage{amssymb}
\usepackage{amsthm}
\usepackage{bm}
\usepackage{dcolumn}
\usepackage{epsfig}
\usepackage{graphicx}
\usepackage{graphics}
\usepackage[latin1]{inputenc}
\usepackage{latexsym}
\usepackage{rotating}
\usepackage{comment}
\usepackage{xspace}

\usepackage[dvipsnames, usenames]{xcolor}
\definecolor{linkcolor}{rgb}{0.0,0.3,0.5}
\usepackage[hypertexnames=false, unicode, colorlinks=true, linkcolor=linkcolor,
citecolor=linkcolor, filecolor=linkcolor,urlcolor=linkcolor,
pdfusetitle]{hyperref}

\newcommand{\model}{{\em EMRISur1dq1e4}\xspace}

\usepackage[normalem]{ulem} 

\begin{document}

\title{A Surrogate Model for Gravitational Wave Signals from Comparable- to Large- Mass-Ratio Black Hole Binaries}

\author{Nur E. M. Rifat} \affiliation{Department of Physics and Center for Scientific Computing \& Visualization Research, University of Massachusetts, Dartmouth, MA 02747}

\author{Scott E. Field} \affiliation{Department of Mathematics and Center for Scientific Computing \& Visualization Research, University of Massachusetts, Dartmouth, MA 02747}

\author{Gaurav Khanna} \affiliation{Department of Physics and Center for Scientific Computing \& Visualization Research, University of Massachusetts, Dartmouth, MA 02747}

\author{Vijay Varma} \affiliation{Theoretical Astrophysics, California Institute of Technology, Pasadena, CA 91125, USA}

\begin{abstract}
Gravitational wave signals from compact astrophysical sources such as those observed by LIGO and Virgo 
require a high-accuracy, theory-based waveform model for the analysis of the recorded signal. 
Current inspiral-merger-ringdown models are calibrated only up to moderate mass ratios, thereby limiting their applicability to signals from high-mass ratio binary systems.
We present \model, a reduced-order surrogate model for gravitational waveforms of $13,500M$ in duration and including several harmonic modes for
non-spinning black hole binary systems with mass-ratios varying from $3$ to $10,000$ thus vastly expanding the parameter range beyond the current models. This surrogate model is trained on waveform data generated by point-particle black hole perturbation theory (ppBHPT) both for large mass-ratio and comparable mass-ratio binaries. 
We observe that the gravitational waveforms generated through a simple application of ppBHPT to the comparable mass-ratio cases agree surprisingly well with those from full numerical relativity after a 
rescaling of the ppBHPT's total mass parameter. 
This observation and the \model surrogate model will enable data analysis studies in the high-mass ratio regime, including potential intermediate mass-ratio signals from LIGO/Virgo and extreme-mass ratio events of interest to the future space-based observatory LISA.
\end{abstract}

\maketitle

\noindent {\em Introduction} -- As the LIGO~\cite{TheLIGOScientific:2014jea} and Virgo~\cite{TheVirgo:2014hva} detectors improve their sensitivity, gravitational wave (GW)
detections~\cite{Abbott:2016blz, TheLIGOScientific:2017qsa, Abbott:2016nmj,
Abbott:2017vtc, Abbott:2017gyy, Abbott:2017oio, LIGOScientific:2018mvr} are
becoming routine~\cite{Aasi:2013wya, LIGOScientific:2018jsj}. In the current observing run, for example,
gravitational-wave events are now being detected multiple times a month~\cite{alerts}.
Among the most important sources for these detectors are binary black hole (BBH) systems, in which two black
holes (BHs) radiate energy through GWs, causing them to inspiral, merge, and finally settle down into a single black hole through a ringdown phase. 

To date all LIGO/Virgo events show support only for systems with mass ratios\footnote{We use the convention $q=m_1/m_2$, where $m_1$ and
$m_2$ are the masses of the component black holes, with $m_1\ge m_2$.} $q = m_1 / m_2 < 8$~\cite{GWTC12019}. Nevertheless, one should expect to observe larger mass ratio systems in the future. For example, the first and second observing runs~\cite{GWTC12019} have already observed compact objects over a mass
range of $1.3 M_{\odot}$ to $85 M_{\odot}$ suggesting combinations involving
mass-ratios in the range of $10$ to $20$ are not unreasonable for LIGO/Virgo to detect, especially, if the lighter member of the binary is a neutron 
star or for BBH systems within the accretion disks of Active Galactic Nuclei~\cite{mckernan2019montecarlo}.
A third generation (3G) ground-based detector, like
the Einstein Telescope or Cosmic Explorer~\cite{hild2011sensitivity,kalogera2019deeper,vitale2017parameter}, may be able to reach up to redshifts beyond 10 and with an improved low-frequency sensitivity limit, implying an increased 
rate of detection of BBH events with unequal mass ratios~\cite{Amaro_Seoane_2018,reitze2019cosmic}. 
Intermediate mass ratio inspirals are also one of the key target sources of the future LISA space-based gravitational wave~\cite{Miller_2009,hughes2006brief,amaro2007intermediate}
detector along with extreme mass ratio systems comprised of a small compact body (possibly a neutron star or stellar mass black hole) orbiting a supermassive black hole (at a galactic center)~\cite{hughes2006brief,amaro2017laser,gair2004event,amaro2007intermediate}. 

In all of these cases we need accurate and fast-to-evaluate inspiral-merger-ringdown (IMR) models covering a range of large- to extreme- mass ratio systems. Such models are needed to maximize the science output of data collected by 
ground-based detectors or to perform mock data analysis studies for LISA and 3G detectors.

Successful detection and parameter estimation relies on being able to compute, from accurate numerical relativity (NR) simulations, the detailed {\em waveform} signal template for such systems. Because solving the Einstein field equations for thousands to millions of potential astrophysical sources is exceedingly challenging, several approximate 
waveform models that are much faster to evaluate have been developed~\cite{Khan:2018fmp, Cotesta:2018fcv, London:2017bcn, Pan:2013rra, Bohe:2016gbl, Khan:2015jqa, Hannam:2013oca,Taracchini:2013rva, Pan:2011gk, Mehta:2017jpq, Babak:2016tgq, damour2013improved, bernuzzi2010binary, bernuzzi2011binary, bernuzzi2011binary2, nagar2018time},
including an effective one body model~\cite{taracchini2013modeling,taracchini2014small} 
calibrated up to $q=100$ using results from black hole perturbation theory ~\cite{Bohe:2016gbl}.
These models assume an
underlying phenomenology based on physical considerations, and calibrate any
remaining free parameters to NR simulations.
Within the LISA waveform modelling community, the computational expense of perturbation-theory waveforms presents a similar bottleneck. 
By relying on a combination of approximations, progress has been make towards the development of 
``kludge" models which can generate waveforms quickly while capturing the qualitative features of extreme mass ratio inspiral (EMRI) waveforms~\cite{PhysRevD.69.082005,PhysRevD.75.024005,PhysRevD.69.082005,chua2017augmented,Chua_2015}.

Surrogate modeling~\cite{Field:2013cfa,purrer2014frequency} is an alternative
approach that doesn't assume an underlying phenomenology and has been applied
to a diverse range of problems~\cite{Field:2013cfa, purrer2014frequency,
    Lackey:2016krb, Doctor:2017csx, Blackman:2015pia, Blackman:2017dfb,
Blackman:2017pcm, Huerta:2017kez, chua2018roman, Canizares:2014fya,
Galley:2016mvy,varma2019surrogate,varma2019surrogate2,Varma:2018aht,chua2018roman,lackey2019surrogate,Bohe:2016gbl,Lackey:2018zvw}. These models follow a data-driven learning strategy, directly
using waveform training data collected by running numerically expensive partial differential equation (PDE) solvers.
Surrogate models are accurate in the region of parameter space over which they were trained
as well as extremely fast to evaluate. For example, for our model
the underlying solver used to generate a single training waveform takes about 2 hours while its corresponding
surrogate
can be 
evaluated in under a second.

Modeling IMR signal templates for black hole binary systems with moderate to large mass ratios
has remained challenging. One practical reason is that 
comparable-mass binaries, a dominant source for currently
operational ground-based detectors, have received significant attention
from the waveform modeling community.
Furthermore, 
this is a parameter regime that is particularly challenging 
for NR as the small length scales introduced by the smaller BH impose a very high grid resolution requirement.
In the {extreme} cases, i.e. when one black hole is {supermassive} like those at the center of most galaxies, the mass-ratio may approach 
$q \sim 10^9$. 
These are well beyond the scope of NR and are typically well-suited for black hole perturbation theory. 

In this {\em Letter} we present a surrogate model for gravitational waveforms emitted from non-spinning black hole binary systems that span an extremely wide range of mass-ratios, from $q = 3$ to $q=10,000$. 
This is the first surrogate model
that covers such a wide range of mass-ratios. The model includes all of the phases of the system's evolution starting from a slow inspiral 
through plunge and ringdown, and includes not only the dominant quadrupole mode, but also several of the most important higher harmonic modes that are especially important at 
larger mass ratio~\cite{Varma:2016dnf,bustillo2018tracking,2018PhRvD..98b4019P,Littenberg:2012uj,kalaghatgi2019parameter}. The model 
spans $13,500M$ in duration, which for a $q=10$ and $q=10^4$ system corresponds to $32$ and $144$ orbital cycles, respectively.
This model can be immediately used in data analysis studies or phenomenological model building efforts that involve large-mass ratio systems, and serves as a proof-of-principle that the surrogate modeling methodology developed for LIGO-type sources remain applicable for LISA-type sources. 
In future work we will extend our model to include spinning BHs and more orbits.

The {\em training} data we use to build this reduced-order surrogate model is generated using the point-particle black hole perturbation theory (ppBHPT) framework, 
i.e. a high-performance
{\em Teukolsky equation}~\cite{teukolsky1973perturbations}
solver code (using a point-particle source) in the 
time-domain~\cite{sundararajan2007towards, sundararajan2008towards, sundararajan2010binary, zenginouglu2011null}. 
While black hole perturbation theory's domain of validity is typically
taken to be 
very high mass-ratio binaries, 
it is interesting to note that a simple rescaling of the mass parameter is sufficient to 
achieve
accurate agreement with NR waveforms for mass-ratios less than $10$. 
That perturbation-theory waveforms agree at all with NR for small-mass ratio systems is somewhat remarkable given that this regime is typically considered beyond perturbation theory's domain of validity. 

\noindent {\em Background on ppBHPT} -- 
In the context of the large mass-ratio limit of a black hole binary system, the system's dynamics can be 
described
using Kerr black hole perturbation theory. In this approach, the smaller black hole is modeled as a point-particle with no internal structure, moving in the space-time of the larger Kerr black hole. Gravitational radiation is computed by evolving the perturbations generated by 
solving the Teukolsky master equation with a
particle-source~\cite{sundararajan2007towards, sundararajan2008towards, sundararajan2010binary, zenginouglu2011null}.

We implemented this ppBHPT approach in two steps. First, we compute the trajectory taken by the point-particle, and then we use that trajectory to compute the gravitational wave emission. For the first step, the particle's motion can be 
characterized by three distinct regimes
-- an initial adiabatic inspiral, in which the particle 
follows a sequence of 
geodesic orbits, driven by radiative energy and angular momentum losses computed 
by solving the frequency-domain Teukolsky equation~\cite{10.1143/PTP.112.415,10.1143/PTP.113.1165,10.1143/PTP.95.1079,ThorweThesis} with an 
open-source code GremlinEq~\cite{gremlin,osullivan2014strongfield,Drasco_2006}; a late-stage geodesic plunge into the horizon; and a transition regime between those two~\cite{ori2000transition,hughes2019learning,sundararajan2010binary,apte2019exciting}
For the low mass-ratio cases, unsurprisingly, the Ori-Thorne transition trajectory algorithm doesn't perform very well. This results in a small jump in the velocities of the point-particle as it exits the adiabatic inspiral and also when it begins the plunge. This jump results in some small unphysical oscillations in the waveforms, especially in some of the higher-order modes. We correct for this by using a ``smoothening'' procedure. 
It should be noted that our trajectory model does not include the effects of the conservative or second-order self-force~\cite{Hinderer_2008}, although once these post-adiabatic corrections are known (see, for example, Refs.~\cite{Gralla_2012,Pound_2012,pound2019secondorder}) they could be easily incorporated to improve the accuracy of the inspiral's phase. 

\begin{figure}[bh]
\includegraphics[width = 0.45\textwidth]{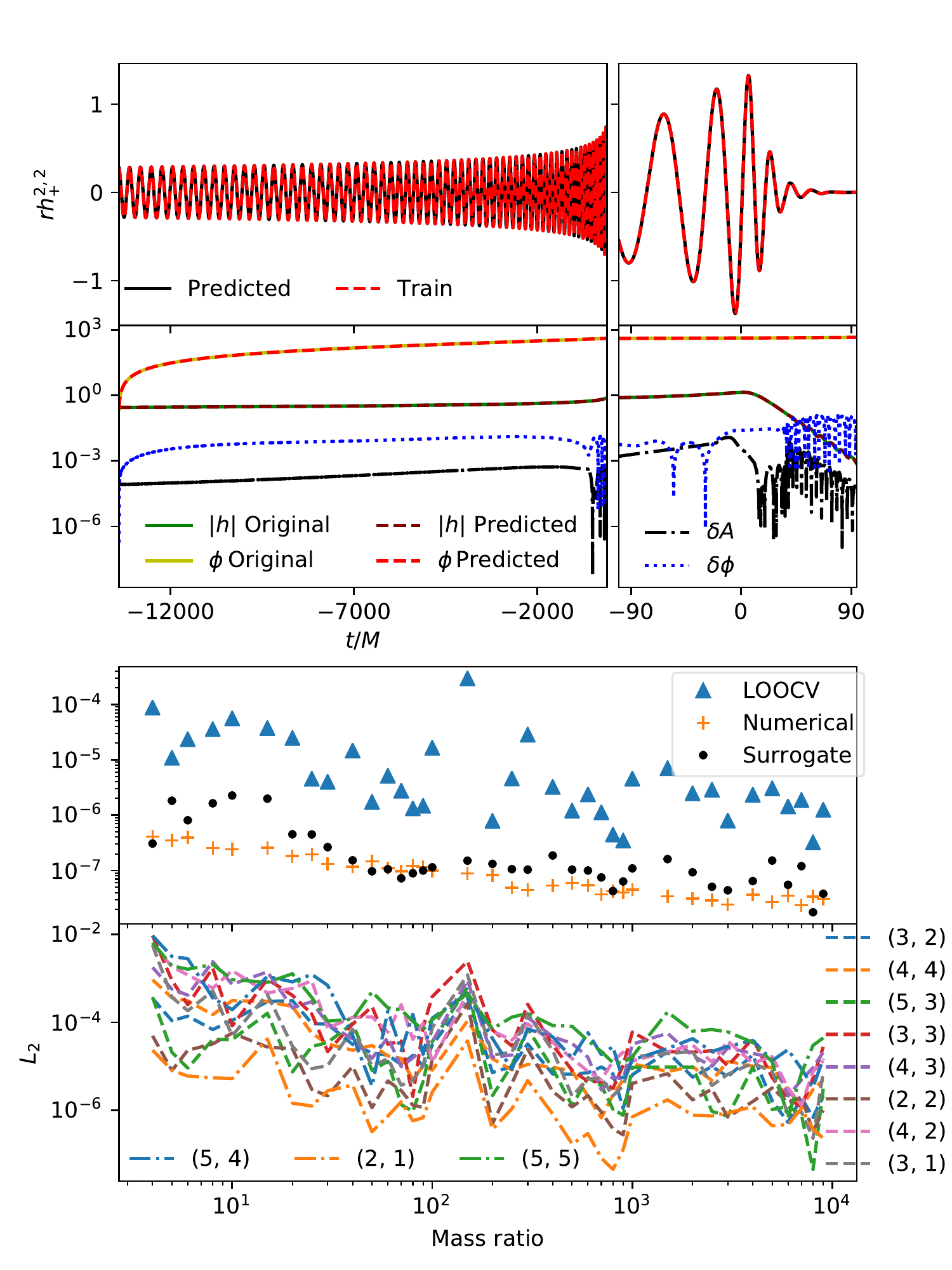}
\caption{
The {\bf bottom subfigure} depicts each mode's relative error by comparing the surrogate model and the training data. The relative error is computed from Eq.~\eqref{eq:alpha_22} with $\alpha=1$ and using the appropriate mode in place of where it says ``$h^{22}"$. The {\bf penultimate subfigure} shows the numerical truncation errors (orange plus) estimate the quality of
the training waveform data by comparing two numerical simulations of increasing resolution. We also compared the full surrogate (including all $11$ harmonic modes) and the training data (black circles) and a leave-one-out cross-validation (LOOCV) trial surrogate and the training data (blue triangles). The largest LOOCV error is for q = 4, for which the $h_+$ polarization's quadrupole mode is shown in the {\bf top subfigure}. The {\bf second subfigure} reports on the error in the amplitude and phase for this case; our full surrogate, trained on the entire data set of $37$ waveforms, is more accurate than the LOOCV diagnostics shown.
}
\label{fig:all_errors}
\end{figure}

With the trajectory of the perturbing compact body fully specified, we then solve
the inhomogeneous Teukolsky equation in the time-domain while feeding the trajectory information from the first step into the particle source-term of the equation.
In particular, (i) we first rewrite the Teukolsky equation using compactified hyperboloidal coordinates that allow us to extract the gravitational waveform directly at null infinity while also solving the issue of unphysical reflections from the artificial boundary of the finite computational domain; (ii) we take advantage of axisymmetry of the background Kerr space-time, and separate the dependence on azimuthal coordinate, thus obtaining a set of (2+1) dimensional PDEs; (iii) we then recast these equations into a first-order, hyperbolic PDE system; and in the last step (iv) we implement a two-step, second-order Lax-Wendroff, time-explicit, finite-difference numerical evolution scheme. The particle-source term on the right-hand-side of the Teukolsky equation requires some specialized techniques for such a finite-difference numerical implementation and, for technical reasons, we set the spin of the central black hole ($a = 10^{-6}$) sufficiently close to zero.
Additional details can be found in our earlier 
work~\cite{sundararajan2007towards, sundararajan2008towards, sundararajan2010binary, zenginouglu2011null}
and the associated references. Our numerical evolution scheme
is implemented using OpenCL/CUDA-based GPGPU-computing which allows for
very long duration and high-accuracy computations within a reasonable time-frame. Numerical errors in the phase and amplitude are typically on the scale of a small fraction of a
percent~\cite{mckennon2012high} (cf. Fig.~\ref{fig:all_errors}).

\noindent {\em Description of the Surrogate Model} \model --
Our surrogate model is built using a combination of methodologies proposed in previous 
works~\cite{Blackman:2015pia,Field:2013cfa,purrer2014frequency}, which we briefly summarize here.


We collect our waveform training data by numerically solving the inhomogeneous Teukolsky equation at 37 different values of the mass ratio $q$ (cf. the bottom panel of Fig.~\ref{fig:all_errors}) and for each value of $q$ extract the harmonic modes, $h^{\ell,m}(t;q)$, for 
$(\ell,m) = \{(2, \{2,1\}), (3,\{3,2,1\}), (4,\{4,3,2\}),(5,\{5,4,3\})\}$. 
Following~\cite{Blackman:2015pia}, we enact a time shift and physical rotation about the z-axis such that (i) each waveform's time is shifted such that $t=0$ occurs at the peak of the 
the (2,2) mode's amplitude, $\left|h^{22}\right|$, and (ii) all the modes' phases are aligned by performing a frame rotation about the z-axis such that 
at the start of the waveform 
$\phi^{22} = 0$ and $\phi^{21} \in (-\pi, 0]$, where $\phi^{22}$ and $\phi^{21}$ are the phases of the  complex $h^{2,2}$ and $h^{2,1}$ modes, respectively.
This pre-processing alignment step ensures that all of the training-set waveforms now depend smoothly on the parameter $q$. 

After alignment we decompose the waveform into {\em data pieces} which are simpler to model. In our case, we choose the waveform modes' amplitude and phase as our data pieces, and interpolate these onto a time grid $\left[-13404, 94 \right]$M with $\Delta t = 0.05$M. Following Refs.~\cite{Blackman:2015pia,Field:2013cfa}, we construct an empirical interpolant (EI)~\cite{Maday:2009,chaturantabut2010nonlinear} (an interpolant whose basis and nodes are learned by applying optimization methods to the training set) for 
each data piece;
there are 11 modes provided by the ppBHPT solver and so we construct $22$ empirical interpolants in total (cf. Eq.~3 of Ref.~\cite{Blackman:2015pia}). Note that we model $m>0$ modes only since the negative modes, $h^{\ell, -m} = (-1)^{\ell} h^{\ell,m}{}^*$, are related to the positive modes due to symmetry of the system under reflections about the orbital plane.

The empirical interpolant gives a compact representation for each data piece (and hence the full waveform) in the training set 
by permitting the full time-series to be reconstructed through a significantly sparser sampling 
defined by the EI nodes. To predict new waveforms not in the training set, at each EI node we 
model the data pieces' parametric dependence on $q$ with a spline~\cite{purrer2014frequency}.

\begin{figure}[h]
\includegraphics[width = 0.45\textwidth]{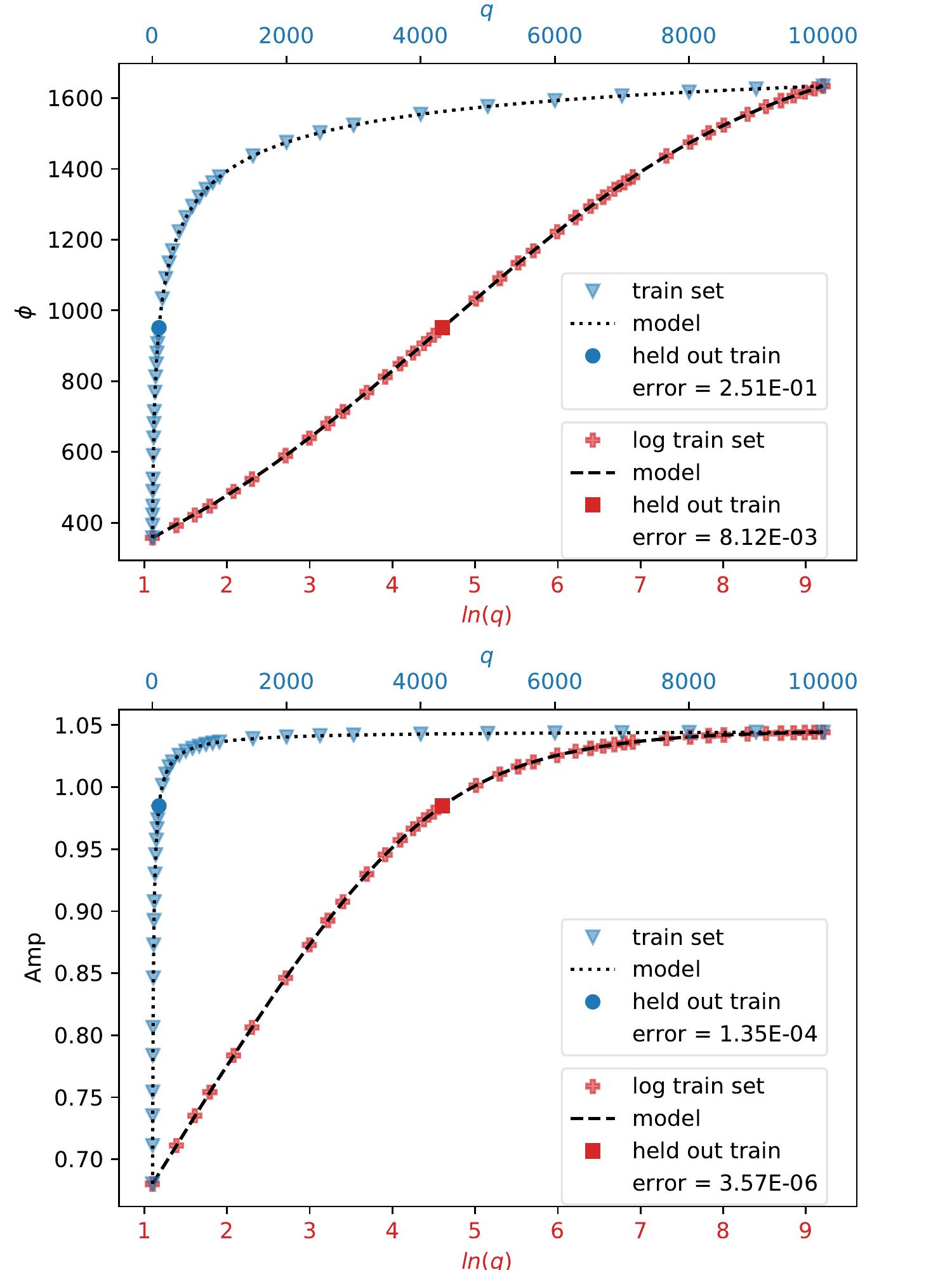}
\caption{At each empirical interpolation node we build surrogate models for 
amplitude, $A^{2,2}(q) \approx A_{\tt S}^{2,2}(q)$, and phase
$\phi^{2,2}(q) \approx \phi_{\tt S}^{2,2}(q)$ across the training region. One of the key methodological improvements pursued here is modeling the training data after performing a logarithmic transformation (red asterisk) of the independent variable $\{\ln(q), A^{2,2}(q)\}$ and $\{\ln(q), \phi^{2,2}(q)\}$. Here we show the LOOCV model error at $q=100$, which leads to nearly two orders of magnitude improvement as compared with no transformation (blue triangle). Data for the other harmonic modes  and empirical interpolation node data show similar improvement.}
\label{fig:ampphase_example}
\end{figure}

Two examples are given in Fig~\ref{fig:ampphase_example}, where we show the training data and model for the $(2,2)$ mode's amplitude and phase at some randomly selected EI node; by fixing the time this data is a function of $q$ only. Our data-piece models are built using degree 2 interpolating splines without any smoothing factors. As shown in Fig.~\ref{fig:ampphase_example} we find significantly better accuracy when modeling the data after performing a logarithmic transformation of the independent variable. This was first used in Refs.~\cite{varma2019surrogate2, Varma:2018aht}, and we suspect this will be important for any model seeking to cover large ranges of the mass ratio. The remaining $10$ subdominant modes follow the same approach.

When evaluating the surrogate waveform, we first evaluate each surrogate waveform data piece
at the requested value of $q$ and use the EI representation to reconstruct the surrogate prediction for the waveform as a dense time-series. The full surrogate, $h_{\tt S}$, can be written as
\begin{align} \label{eq:full}
    h_{\tt S}(t,\theta,\phi;q) = \sum_{\ell,m} h_{\tt S}^{\ell,m}(t;q) {}_{-2}Y_{\ell m} (\theta,\phi) \,,
\end{align}
where ${}_{-2}Y_{\ell m}$ are the spin ($-2$) weighted spherical harmonics and
models for each harmonic mode (a single complex function), $h_{\tt S}^{\ell,m}(t;q) = A_{\tt S}^{\ell,m}(t;q) \exp(- \mathrm{i} \phi_{\tt S}^{\ell,m}(t;q))$, are defined in terms of models of the amplitude and phases (two real functions).

To assess the surrogate model's error, we perform some of the tests described in Ref.~\cite{blackman2017surrogate} using a relative $L_2$-type norm (we compute the norm of the error through a time-domain overlap integral with a white-noise curve) given exactly by Eq.~(21) in Ref.~ \cite{blackman2017surrogate}. This measures the full waveform Eq.~\eqref{eq:full} error over the sphere and automatically includes error contributions from all of the harmonic modes. In Fig.~\ref{fig:all_errors} we (i) check that the surrogate model can reproduce all 37 ppBHPT waveforms used to train the surrogate (black circles), (ii) perform a leave-one-out cross validation study to asses the model's ability to predict new waveforms it was {\em not} trained on (blue triangles), and (iii) compare both errors to the numerical truncation error of the Teukolsky solver used to produce the training data (orange plus). We find that the model errors remain extremely small over the range of mass ratios $q=3$ to $10^4$, although a they are a bit larger than the errors in the training data itself. We remind the reader that these comparisons are between the model, \model, and the output of the Teukolsky solver. Waveforms generated within the ppBHPT framework are expected to become more accurate as $q$ becomes large. Next we provide evidence for using ppBHPT waveforms even at small mass ratios.

\noindent {\em Waveforms from Comparable Mass Binaries using Perturbation Theory} -- 
We now proceed to compare the model output with full NR data. 
This comparison is naturally restricted to low mass-ratios $q\leq 10$.
For the high mass-ratio cases, extensive comparisons 
with EOB have been performed previously in the context of the EMRI data itself~\cite{barausse2012}, so we do not focus on those cases. Additionally, there is a
lack of models and data for the intermediate ranges, say from $q=10$ to $q=10^4$, so we leave that domain open for future comparisons.

\begin{figure}[h]
\includegraphics[width = 0.45\textwidth]{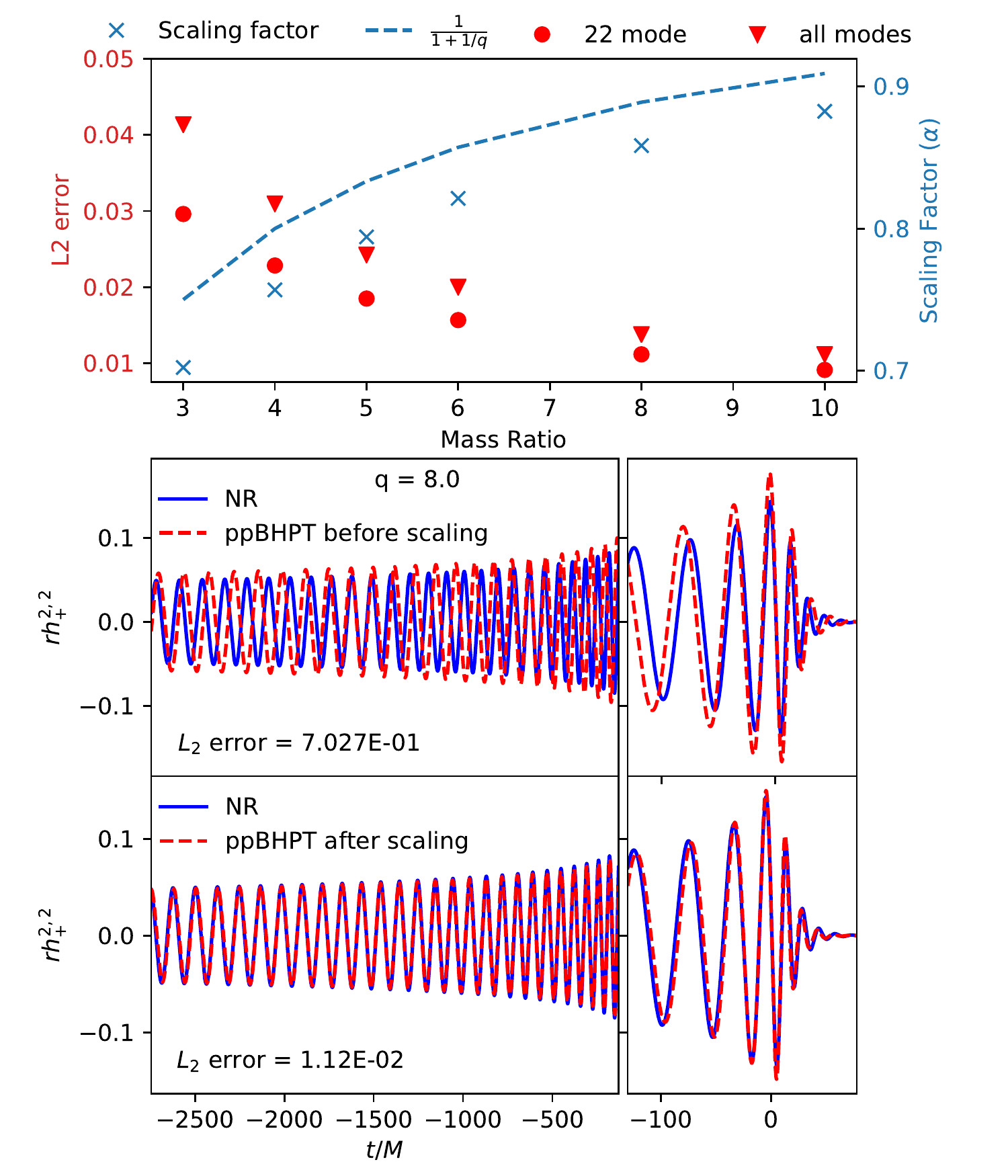}
\caption{Waveform difference between numerical relativity and ppBHPT waveforms before and after rescaling the ppBHPT's total mass parameter. {\bf Bottom panel}: These two figures focus on the case $q=8$. Before rescaling the NR (solid blue) and ppBHPT (dashed red) waveforms are noticeably different in both their amplitude and phasing and have an L2 difference of $7.03 \times 10^{-1}$. We find the optimum value of the rescaling parameter to be $\alpha = 0.85837$
and modify the ppBHPT waveform according to Eq.~\eqref{eq:EMRI_rescale}. After rescaling the NR and modified ppBHPT waveforms demonstrate remarkable agreement with one another. {\bf Top panel}: We repeat this comparison procedure for mass ratios $3 \leq q \leq 10$ where NR data is available, and compute both the optimal scaling factor and the L2 difference between waveforms (NR vs scaled ppBHPT) for each case. In all cases the differences before rescaling 
are order unity, while the agreement between the rescaled ppBHPT and NR waveforms is $\approx 1$\%. The dotted line refers to a naive value of 
$\alpha = 1/(1+1/q)$ set by including the mass the of smaller black hole as part of the background spacetime.
}
\label{fig:nr_comparison}
\end{figure}

One complication that appears when we attempt to perform a careful comparison with 
NR 
is how to set an overall mass-scale for the comparison and, more generally, 
identify parameters.
Indeed, all dimensioned quantities in both ppBHPT and NR frameworks are written in terms of a freely-specifiable mass-scale. For ppBHPT this scale is selected to be the background black hole spacetime's mass parameter, while the sum of the Christodoulou masses of each black hole is the choice implemented in the NR code~\cite{Boyle:2007ft,boyle2019sxs}. If the background black hole's mass is
set to $1$, naively we might expect the corresponding
NR simulation's total mass (its mass-scale) to be $1 + 1/q$. This 
straightforward identification 
works well
when comparing post-Newtonian and NR waveforms~\cite{Boyle:2007ft},
while only in the limit of large $q$ does the ppBHPT mass-scale
seem to approach the naive one.

To address this uncertainty,
we perform a rescaling of our surrogate model data ($t \to \alpha t$, $r \to \alpha r$) using a {\em single} parameter, $\alpha$, which, due to coordinate invariance of GR, describes a physically equivalent solution.
This simultaneous rescaling of $r$ and $t$ 
may also be interpreted as keeping the coordinates fixed while modifying the total mass parameter as $M \rightarrow M / \alpha$.
In particular, we propose modifying the ppBHPT surrogate model presented above according to the formula
\begin{align} \label{eq:EMRI_rescale}
h^{\ell,m}_{\tt S, \alpha}(t ; q)= {\alpha} h^{\ell,m}_{\tt S}\left( t \alpha;q \right) \,,
\end{align}
where $\alpha$ is set by minimizing the difference
\begin{align} \label{eq:alpha_22}
    \min_{\alpha} \frac{\int \left| h^{22}_{\tt S, \alpha}(t ; q) - h^{22}_{NR}(t ; q) \right|^2 dt}{\int \left|  h^{22}_{NR}(t ; q) \right|^2 dt} \,,
\end{align}
between our model and a handful of 
nonspinning NR waveform datasets~\cite{SXSCatalog,Mroue:2013xna,boyle2019sxs}
for the $(2,2)$ harmonic mode. 
We then fit 
$\alpha(\nu)$
\begin{align}
\begin{split}
\alpha(\nu) =  1 & - 1.352854 \nu - 1.223006 \nu^2 \\ 
& + 8.601968 \nu^3 - 46.74562 \nu^4 
\end{split}
\end{align}
to a polynomial in $\nu$, which is the symmetric mass $\nu = q / (1+q)^2$.
Details of this parameter are presented in Fig.~\ref{fig:nr_comparison} alongside the error (computed as Eq.~\eqref{eq:alpha_22})
between the rescaled surrogate model and NR waveforms. As expected the rescaling parameter approaches unity as the mass-ratio increases,
and the error
decreases according to the trend 
${\epsilon}(\nu) = 0.082111353 \nu + 0.2698017 \nu^2 +  0.7116969 \nu^3$.
We conjecture that these fitting formula will continue to 
be applicable for values $q > 10$.
To test our conjecture, we compare our model against a new $q=15$ NR simulation performed using the SpEC code with recent algorithmic improvements~\cite{highq1, highq2}. We find the (2,2) modes agree to $6.1 \times 10^{-3}$, which is consistent with our predicted accuracy formula's value of  $5.8 \times 10^{-3}$.

\begin{figure}[h]
\includegraphics[width = 0.45\textwidth]{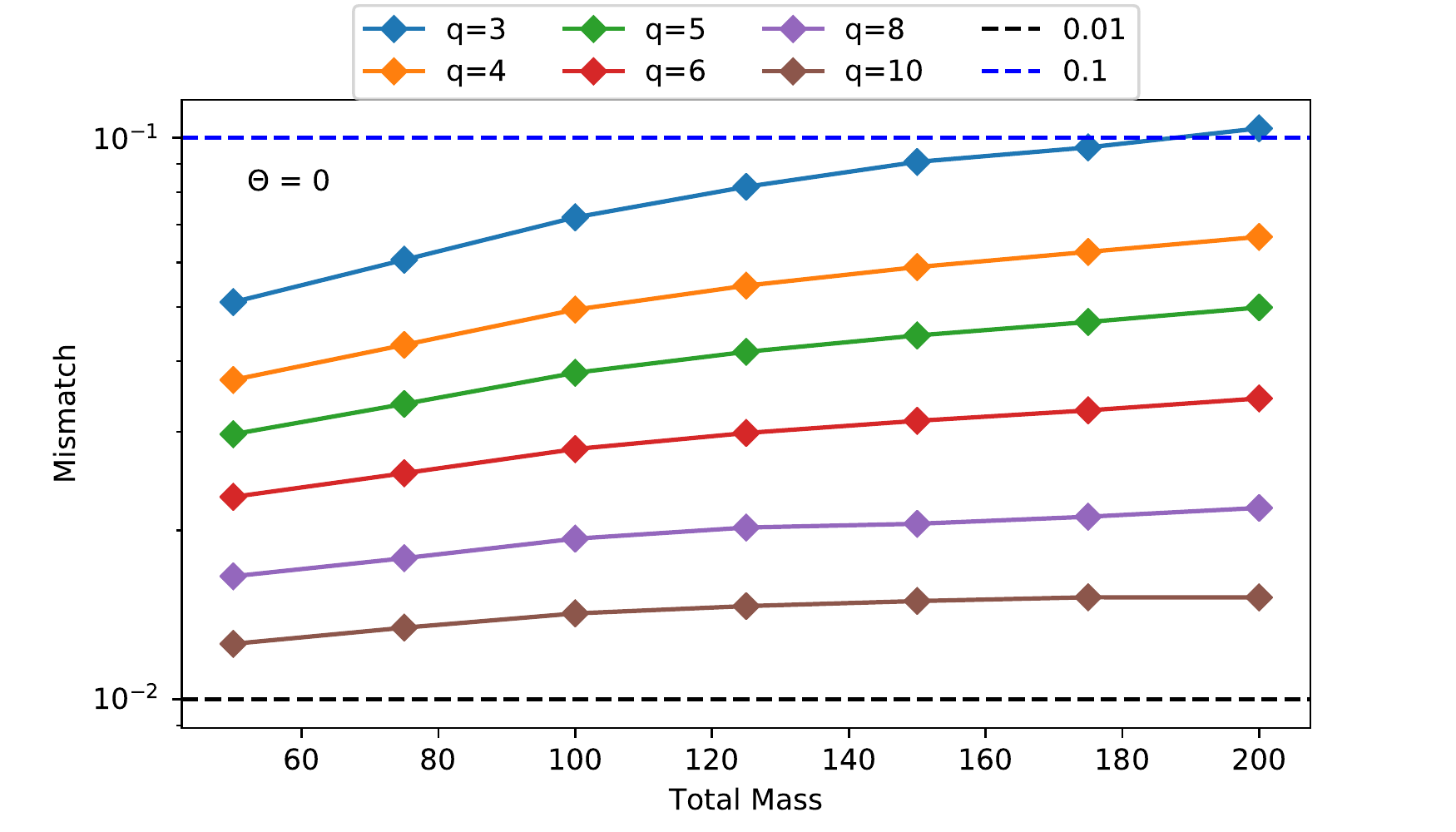}
\caption{Frequency-domain mismatch between NR and rescaled ppBHPT waveforms using all available modes. The mismatch computation uses an advanced LIGO design sensitivity curve~\cite{aLIGODesignNoiseCurve}, an upper frequency of $4096$ Hz, and a variable lower
frequency set to the (2,2) mode's initial instantaneous frequency. We show a result for an inclination angle of 0, while non-zero inclinations are typically a factor of $\sim 1.5$ times larger. As expected, the ppBHPT and NR waveforms show better agreement at larger mass ratios. }
\label{fig:mismatches}
\end{figure}

Note that we use precisely the same $\alpha$ parameter to enact an analogous rescaling for all the higher-modes too. Since $\alpha$ has been optimized using the (2,2) mode  data, the subdominant modes do not achieve relative errors as low as the (2,2) mode. Nevertheless, these higher-modes are 
still well modeled and the overall error, including error contributions from all modes, is nearly the same as the (2,2)-mode only error (cf.~Fig.~\ref{fig:nr_comparison}).

As a final test, in Fig.~\ref{fig:mismatches} we show a noise-weighted mismatch (cf. Eq.~(22) of Ref.~\cite{Blackman:2017dfb}) between NR and the ppBHPT waveforms using all available modes. We continue to find good agreement and, as expected, as the mass ratio increases the mismatch decreases. Finally, the mismatch between our model and the new $q=15$ NR simulation (not shown) is around $.01$ for the range of total masses considered.

\noindent {\em Summary} -- In this {\em Letter} we present the first surrogate model, \model, for gravitational wave signals (including higher-order modes) from black hole binary systems over a wide range of
mass-ratios. \model can be used to extend the banks of signal templates for LIGO/Virgo data analysis into larger mass-ratios, and also serve as a useful tool for mock data analyses for future observatories. This model
is publicly available as part of both the Black Hole Perturbation
Toolkit~\cite{BHPToolkit} and GWSurrogate~\cite{gwsurrogate}.
Future work should include obvious extensions to the model such as spin, effects of eccentricity, and spin-orbit precession.

We also perform the first comparison between ppBHPT and NR waveforms, and find that after a rescaling of the ppBHPT's total mass parameter there is surprisingly remarkable agreement even in the comparable mass-ratio regime. Note that the ppBHPT calculation does not incorporate any aspect of the 
dynamics of the background geometry within which the waves travel and nonlinearities beyond radiative corrections to the orbit. This study, which may 
offer some insight into the dynamics of a black hole binary system itself, is part of a growing body of evidence, initiated by Le Tiec 
et al.~\cite{le2011periastron} (see also Refs.~\cite{price2011systematics,price2013black,price2016black}), that suggests perturbation theory with self-force corrections are applicable to nearly equal mass systems~\cite{lewis2017fundamental,zimmerman2016redshift,le2011periastron,le2012gravitational,le2014overlap,le2013periastron} despite there being 
no a priori reason to expect this should be the case.  
As a practical matter, our results suggest that perturbation theory with 
(post-)adiabatic orbital corrections may be used to generate accurate late inspiral, merger, and ringdown 
waveforms in the $q>10$ regime that is especially challenging for NR.

\noindent{\em Acknowledgments} -- We would like to thank Alessandra Buonanno, Scott A. Hughes, Rahul Kashyap, Steve Liebling, Richard Price, Michael P{\"u}rrer, Niels Warburton, and Anil Zenginoglu for helpful feedback on this manuscript. 
We also thank Tousif Islam for significant assistance with porting the surrogate 
model to the Black Hole Perturbation Toolkit, and Matt Giesler and  Mark Scheel for
allowing us to use their new $q=15$ NR simulation in our study.
N.\ R.\ and G.\ K.\ acknowledge research support from 
NSF Grants PHY-1701284 \& DMS-1912716, and ONR/DURIP Grant No.\ N00014181255. S.\ E.\ F.\ is partially supported by NSF grant PHY-1806665. We also thank Scott A. Hughes for his open-source code GremlinEq, which is part of the Black Hole Perturbation Toolkit~\cite{gremlin}. This code enabled us to compute 
the decaying trajectories using the energy-balance approach. 


\section*{References}

\bibliography{References}

\end{document}